\documentclass[aps,prl,twocolumn,groupedaddress]{revtex4}

\usepackage{amsmath}
\usepackage{epsfig}         
\usepackage{color}

\begin{document}



\title{Controlling spontaneous emission with plasmonic optical patch antennas}

\author{C. Belacel$^{1,2,3}$, B. Habert$^4$, F. Bigourdan$^4$, F. Marquier$^4$, J-P. Hugonin$^4$, S. Michaelis de Vasconcellos$^1$, X. Lafosse$^1$, L. Coolen$^{2,3}$, C. Schwob$^{2,3}$, C. Javaux$^5$, B. Dubertret$^5$, J-J. Greffet$^4$, P. Senellart$^1$ and A. Maitre$^{2,3}$}
\affiliation{$^1$Laboratoire de Photonique et de Nanostructures, CNRS, UPR20, Route de Nozay, 91460 Marcoussis, France\\
$^2$Universit\'e Pierre et Marie Curie-Paris 6, UMR 7588, INSP, Campus Boucicaut, 140 rue de Lourmel, Paris, F-75015 France\\
$^3$CNRS, UMR7588, INSP, Paris, F-75015 France\\
$^4$Laboratoire Charles Fabry, Institut d Optique, CNRS, Univ Paris-Sud, Campus Polytechnique, RD128, 91127 Palaiseau cedex\\
$^5$Laboratoire de Physique et d'\'Etude des Mat\'eriaux, CNRS UMR8213, ESPCI, 10 rue Vauquelin, F-75231 Paris, France}



\date{\today}

\begin{abstract}
We experimentally demonstrate the control of the spontaneous emission rate and the radiation pattern of colloidal quantum dots deterministically positioned in a plasmonic patch antenna. The antenna consists of a thin gold microdisk 30 nm above a thick gold layer. The emitters are shown to radiate through the entire patch antenna in a highly directional and vertical radiation pattern. Strong acceleration of spontaneous emission is observed, depending of the antenna size. Considering the double dipole structure of the emitters, this corresponds to a Purcell factor up to 80 for dipoles perpendicular to the disk.
\end{abstract}


\maketitle


Recent years have seen impressive progress in the fabrication of bright sources of quantum light by inserting single quantum emitters in photonic structures\cite{Pelton2002,Dousse2010,Claudon2010,Lee2011}. In dielectric cavities like micropillars\cite{Pelton2002}, microdisks\cite{Kiraz2001} or defects in photonic crystals\cite{Englund2005}, the electromagnetic field is confined on the wavelength scale. In the case of a single mode cavity, a quantum emitter spatially located at the maximum of the optical field experiences an acceleration of spontaneous emission given by the Purcell factor $F_p$, proportional to Q/V where Q is the mode quality factor and V is the mode effective volume\cite{Purcell}. In dielectric cavities, Q  larger than $10^3$ are used to obtain large Purcell factors, ensuring a high coupling to the mode $\beta=F_p/(F_p+1)$ \cite{Gerard1999}.
Although well suited to extract quasi-monochromatic single photons generated by emitters like epitaxial semiconductor quantum dots (QDs) \cite{Pelton2002,Dousse2010}, this approach is not appropriate for spectrally broad single photon emitters operating at room temperature like N-V centers in diamonds\cite{Kurtsiefer2000} or colloidal quantum dots\cite{Michler2000}. Dielectric antennas have recently been shown to provide  broadband single photon collection\cite{Claudon2010,Lee2011}.

\vspace{0.3cm}
Here, we explore the possibility of extracting single photon emission using plasmonic antennas, which can confine the electromagnetic field on highly subwavelength volumes\cite{Novotny2011}. The quality factor of one mode in a plasmonic structure can be as small as 10 and provide broadband control of spontaneous emission while maintaining large spontaneous emission acceleration, an important point to increase the source operation rate. First works took advantage of hot spots of the optical field in metallic nanoparticules or at the end of metallic tips to demonstrate acceleration of spontaneous emission \cite{Novotny2006, Sandoghdar2006}. More recently, either acceleration of spontaneous emission \cite{Kinkhabwala2009,Benson2009} or directional emission \cite{Curto2010, Kosako2010} have been demonstrated in lithography designed plasmonic antennas. Yet, both properties are needed for efficient single photon extraction.

In the present work, we experimentally demonstrate control of  both the radiation pattern and the emission dynamics of emitters coupled to plasmonic patch antennas.  Small clusters of  colloidal quantum dots (QDs)  are deterministically positioned with 25 nm accuracy in the antenna using an in-situ lithography technique\cite{Dousse2008}. The emitters are shown to radiate through the whole patch antenna surface with a directionnal far field pattern. Time resolved measurements show a strong acceleration of spontaneous emission. By fitting the experimental curves to account for the random orientation of the double dipole of the emitters, we show that this corresponds to a decay rate enhancement up to 80 for dipoles perpendicular to the antenna.

\begin{figure}[t]
 \includegraphics[width=8.5cm]{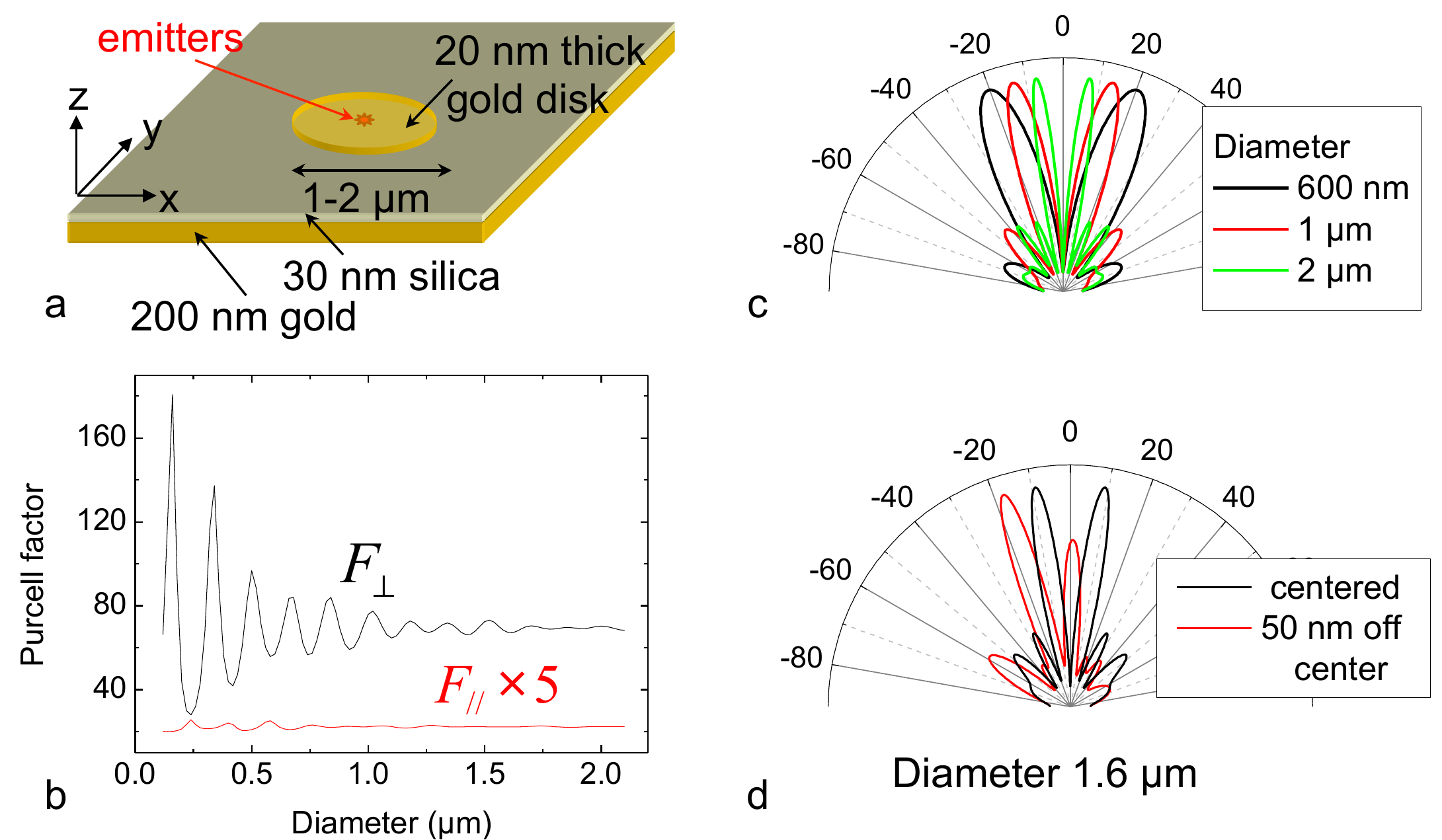}%
 \caption{Plasmonic patch optical antenna. a: Sketch of the patch-antenna structure. b. Calculated Purcell factor for a point emitter centered in a patch antenna with dipole along $z$ (black lines) or along $xy$ (red lines). c,d: Calculated radiation patterns of a point emitter with dipole along $z$ . c:  Calculation for a spatially centered emitter in patch antennas with various diameters. d: Calculation for an antenna with 1.6~$\mathrm{\mu}$m diameter and two positions for the emitter. Black line: emitter perfectly centered in the antenna, red line: the emitter is 50~nm off centered.\label{Fig1}}
 \end{figure}

 Plasmonic patch antenna were recently proposed as a highly promising system for efficient single photon sources\cite{Esteban2010}. The  antenna  structure  is illustrated in figure~\ref{Fig1}(a)). It consists of a thin metallic microdisk separated from a metallic plane by a  few tens of nanometers thick dielectric layer. Coupling of surface plasmons at both dielectric-metal interfaces as well as reflections of surface plasmons at the disk edges results in strongly confined optical modes below the disk area. Esteban and coworkers\cite{Esteban2010} calculated that an emitter inserted in such a structure should experience spontaneous emission enhancements as large as a few hundreds and highly directional emission.

Here, we insert colloidal QDs emitting at  630 nm  in gold plasmonic patch antenna. A 30 nm thick silica layer with refractive index 1.5 is used as dielectric spacer between the two gold layers.  
To describe the device under investigation, we use the Rigorous Coupled
Wave Analysis as implemented in ref.\cite{Armaroli2008}. The field is expanded
over a basis of modes $e^{iKz}f(kr)e^{iL\theta}$ characterized by the
wavevector along the disk normal $K$, the integer $L$ and the complex
eigenvalue $k$ of the radial transverse wavevector; $f(kr)$ is an incoming
or outgoing cylindrical mode. The method is generalized to the study of
non-periodic objects by introducing Perfectly Matched Layers (PML)
\cite{Lecamp2007} at two planes above and below the antenna. The mode
amplitudes are obtained by enforcing the boundary conditions at the
interfaces. We do not include the $\theta$ dependence because the vertical dipole is located at the center of the disk. Note that a monochromatic dipole source in the patch antenna is coupled to a large number of modes with quality factors, on the order of 10 \cite{Esteban2010}. As a result, the Purcell factor, cannot be defined using a single mode quality factor and effective volume. Hereafter, we define  the Purcell factor as  the decay rate of the quantum emitter in the plasmonic antenna normalized by its decay rate in silica.

As discussed in detail in the supplementary material, the quantum emitter decays through two channels:  spontaneous emission of an electromagnetic mode (photon or plasmonic antenna mode) or  direct quenching. A large Purcell factor is useful only if it is not dominated by quenching. The quenching effect is related to the short range non-radiative energy transfer between the quantum emitter and the metallic interface\cite{Ford1984}. In the case of the patch antenna, the 15~nm distance between the quantum emitter and both gold layers has been chosen to ensure that direct quenching effects are negligible \cite{suppmat}. Note that after excitation, the plasmonic antenna mode also decays through two channels: photon emission or Joule losses in the antenna so that a radiative efficiency can be defined \cite{Esteban2010}. The Purcell factor $F$ is thus a direct measure of the enhanced spontaneous emission into plasmonic antenna modes.

Figure~\ref{Fig1}(b) presents the theoretically expected Purcell factor for a single quantum emitter emitting at 630~nm as a function of the patch antenna disk diameter. The modification of spontaneous emission is calculated using a point emitter with a dipole along either $z$ or $x$ direction precisely positioned at the disk center and at equal distance along $z$ between the two metallic layers. An oscillatory behavior is predicted for the Purcell factor $F$ (black curve), corresponding to resonances in the disk which behaves as a cavity for surface plasmons in the $xy$ plane. For a dipole oriented along $z$, the maximum Purcell values $F_{\bot}$ range from 70-80 for micron size antennas and reach a peak value of 180 for 150~nm diameter antennas. For an emitter presenting a dipole in the $xy$ plane, the coupling to the antenna is much less efficient and Purcell factors $F_{/\!/}$ between 4 and 5 are expected for the whole diameter range.

Figure~\ref{Fig1}(c) presents the radiation patterns calculated for various disks diameters. For perfectly centered point emitters with dipole along $z$, the radiation pattern presents several highly directional emission lobes. The larger the disk, the narrower and closer to the z axis the radiation pattern is. A 50~nm off-centering of the emitter position with respect to the disk center leads to a tilt of this radiation pattern, as shown in figure~\ref{Fig1}(d) for an antenna with 1.6 $\mathrm{\mu}$m diameter.

 \begin{figure*}[t]
 \includegraphics[width=13.5cm]{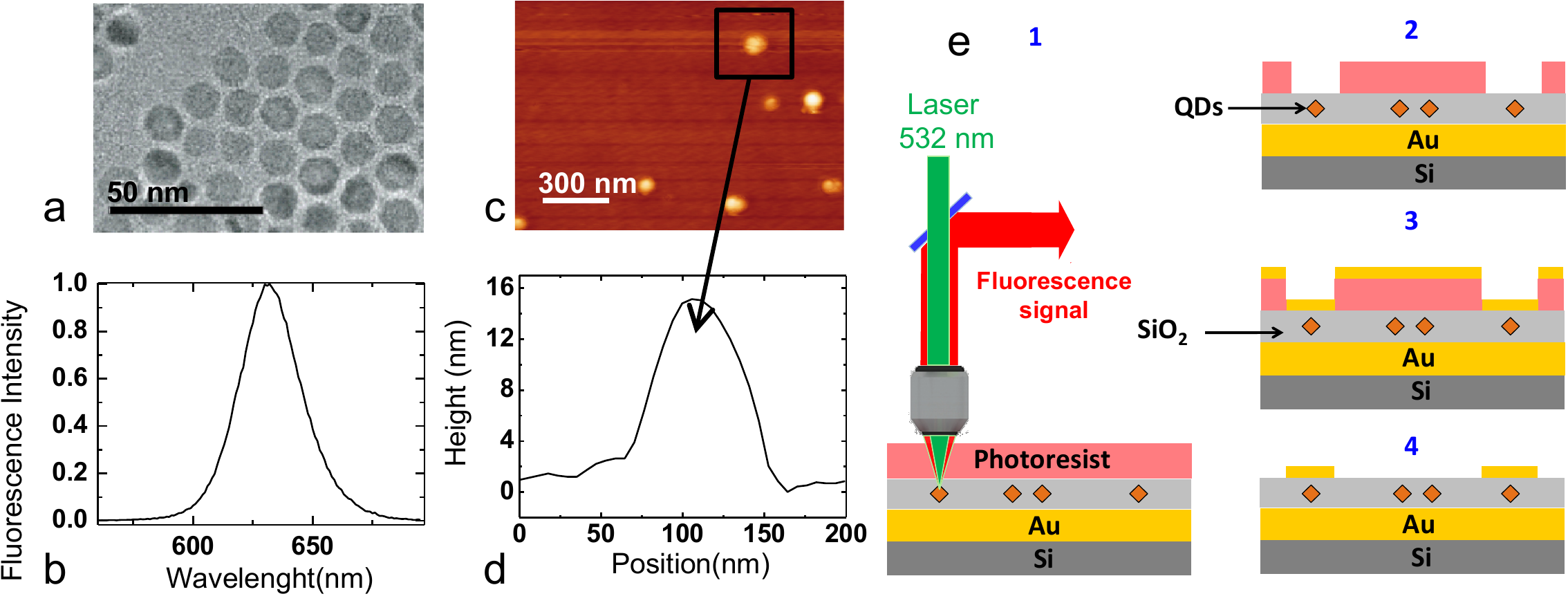}%
 \caption{a: Transmission electron microscopy image of the colloidal core shell quantum dots. The CdSe core diameter is  3 nm and the CdS shell is 5nm thick. The average overall size is around 13 nm. b: Emission spectrum of a single QDs. c: Atomic Force Microscopy (AFM) image of the clusters dispersed on the sample. d: Typical profile of a single cluster measured with AFM. e: Sketch of the technology used to deterministically position the QDs in the antenna  \label{Fig2}}
 \end{figure*}

The emitters under investigation are core/shell CdSe/CdS QDs (figure~\ref{Fig2}(a)) with a   3~nm diameter core and a 5~nm thick shell, providing a total diameter of 13~nm. Their emission is centered around 630~nm as shown by the fluorescence spectra in figure~\ref{Fig2}(b). They emit single photons and their quantum efficiency is around 40 $\%$  and the blinking has been highly reduced thanks to optimized synthesis conditions\cite{Mahler2008}. Their emission pattern is similar to that of a pair of orthogonally polarized incoherent dipoles \cite{Empedocles1999} perpendicular to the QD c-axis. As shown in figure~\ref{Fig1}(b), the orientation of the QD is a crucial parameter for efficient coupling to patch antenna. Rather than fabricating a very large number of devices to statistically find antennas embedding a well-oriented emitter, we choose to study small clusters of nanocrystals made of approximately 50  randomly oriented QDs. 

\begin{figure}
 \includegraphics[width=8.5cm]{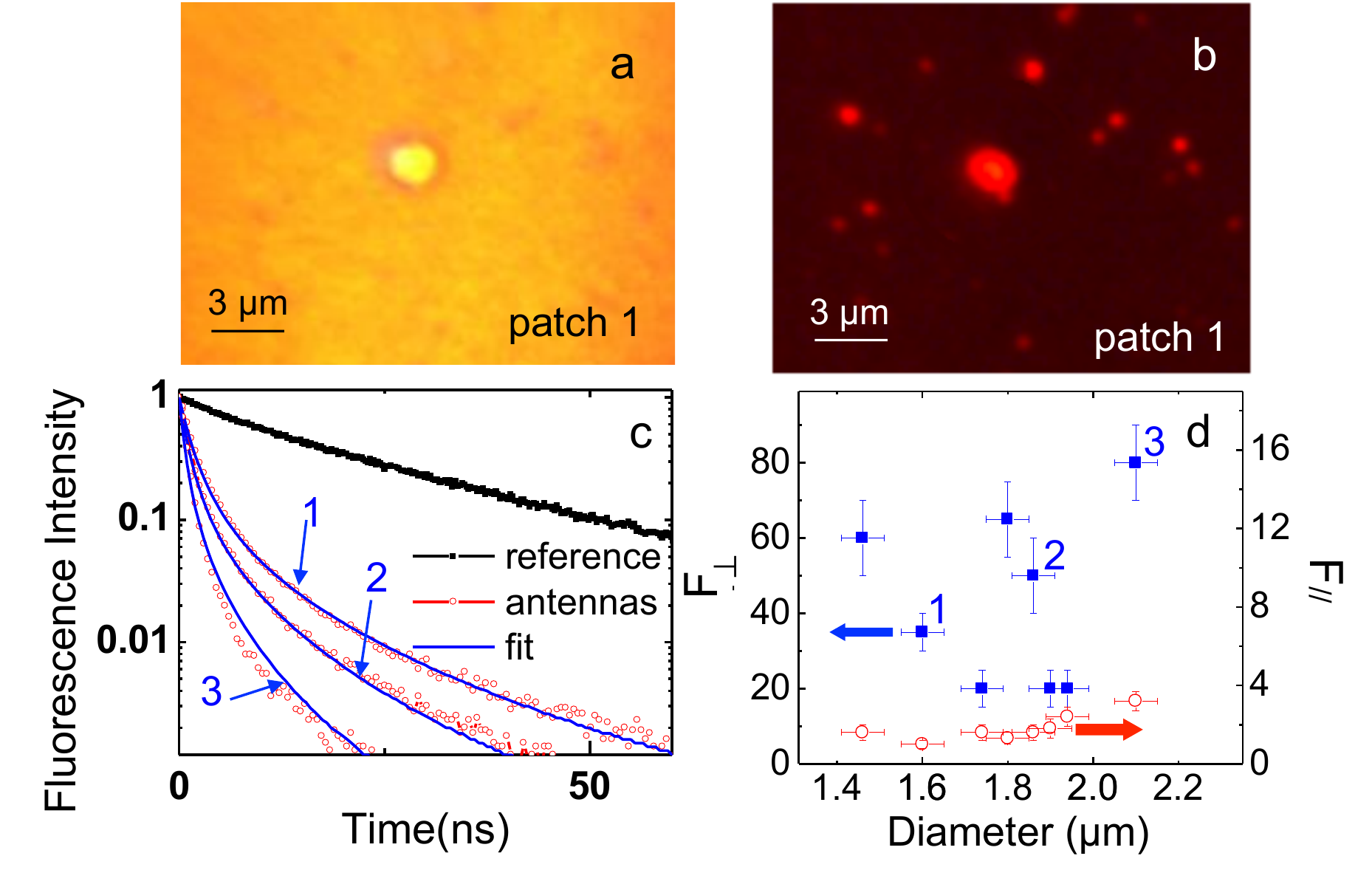}%
 \caption{Controlling the spontaneous emission rate. a: Image of the sample surface under white light illumination. A single gold disk defines the patch antenna (patch 1). b: Fluorescence image of the same sample region under continuous wave optical pumping at 405nm. c: Time dependence of the fluorescence under pulsed 410 nm  excitation. Black symbols: reference fluorescence decay measured for clusters in silica. Red symbols: Fluorescence time dependence for  clusters coupled to various patch antennas (patch 1 to 3). Lines: fits of the decay curves taking into account the Gaussian distribution of decay rates and the random orientation of the QD inside the clusters.  d: Values of $F_{\bot}$ and $F_{/\!/}$ deduced for several antennas. The antenna corresponding to the decays presented in figure 3c are indicated (numbers 1 to 3) \label{Fig3}}
 \end{figure}

The patch antenna sample is fabricated as follows (see figure~\ref{Fig3}(e)). A 200~nm-thick gold layer and a 15~nm silica layer are deposited on a silicon substrate. The QD clusters are formed by diluting an hexane QD solution in isopropanol. The sample is then inserted in the QD clusters solution and pulled out in order to obtain a uniform low density of clusters on the surface. The clusters geometry is  characterized by atomic force microscopy as shown in figure~\ref{Fig2}(c-d). They present a typical thickness around 12-25~nm and lateral size in the $xy$ plane around 70-120~nm.   A second layer of silica (15~nm-thick) is then deposited on the sample followed by a positive photoresist. Following the technique proposed in reference \cite{Dousse2008}, a low-power laser beam at 532~nm focused onto a 1~$\mathrm{\mu}$m spot is used to excite the cluster emission. Moving the cluster in the laser spot on the nanometer scale in order to maximize its fluorescence intensity allows centering the cluster in the laser spot with $\pm$25~nm accuracy (figure~\ref{Fig3}(e), step 1). Then, the excitation power is increased to expose a disk in the resist centered on the QD cluster. The disk diameters can be adjusted by varying the exposure time. After revealing the resist (step 2), a layer of 20~nm of gold is evaporated (step 3) and a lift-off step is performed (step 4). A dozen of antennas are fabricated, with diameters ranging from 1.4 to 2.1~$\mathrm{\mu}$m.

Figure~\ref{Fig3}(a) presents an image of the sample surface obtained with a fluorescence microscope in reflectivity configuration under white light illumination. A 1.6~$\mathrm{\mu}$m diameter gold disk defining a single plasmonic patch antenna is visible (patch 1). The fluorescence image of the same part of the sample is shown in fig.~\ref{Fig3}(b). The emitter is excited with a 405~nm laser and a 510~nm dichroic filter and a 610$\pm$70~nm pass band filter are used to remove signal coming from the excitation light. Small emission spots are visible on the whole surface corresponding to the emission of single QD clusters. The fluorescence spectrum is the same for isolated clusters or clusters under the antenna, showing that the antenna bandwidth is larger than the QD emission spectrum width (30 nm). Quite strikingly, fluorescence is observed over the whole surface of the patch antenna. This is the first signature of the efficient coupling between the QDs and the patch-antenna plasmon modes. Indeed, if the QD emission is strongly accelerated in the antenna optical modes as compared to any other mode, the QD will mainly radiate through these modes.

This interpretation is confirmed by time resolved experiments showing the modification of spontaneous emission of the emitters coupled to the antennas. The QD fluorescence is excited using a 405~nm laser (2.5~MHz repetition rate and 80~ps of pulse width), focused onto a 1 $\mu$m spot with a 0.8 numerical aperture objective. The fluorescence is then collected by the same objective and its time dependence is measured using time correlation of the emission signal with the excitation laser with a 500~ps time resolution.  The black curve in figure~\ref{Fig3}(c)  presents the time dependence of the emission of clusters embedded in a 3D homogeneous silica environment whereas the red curves present the same measurement on clusters coupled to patch antennas. A strong acceleration of the emission dynamics (by a factor 5 to 15) is evidenced for each antenna.

In a homogeneous silica environment, the decay rate of a cluster is given by a Gaussian distribution characterized by a mean decay rate $\Gamma_c$ and a width $w_c $. We then introduce the probability  density 
$\pi_1(\Gamma_Q)=\exp[-(\Gamma_Q-\Gamma_c)^2/2w_c]$
 where $\Gamma_Q$ is the individual QD radiative rate in silica. The values of $\Gamma_c$ and $w_c$ are measured experimentally to be $\Gamma_c= 0,055 \mathrm{ns}^{-1}$ and a width $w_c = 0,020 \pm 0,002 \mathrm{ns}^{-1}$. To account for the experimental decay curves and extract the Purcell factors, we model the coupling of QD clusters to the antenna taking into account the degenerated double dipole character of the QDs and their random orientation.  
For a single QD, with a decay rate $\Gamma_Q$ in silica and  centered under the antenna, with a c-axis oriented with an angle $\theta$ with $z$ , the decay rate scales then as
 $\Gamma=\Gamma_Q/2[F_{\bot}\sin^2\theta+F_{/\!/}(1+\cos^2\theta)]$.
For an isotropic distribution of c-axis orientations, the density of probability of having the c-axis oriented with an angle $\theta$  is given by $\sin \theta$. The conditonnal probability $\pi_2(\Gamma;\Gamma_Q) $ for this specific QD of getting a decay rate $\Gamma$ inside the antenna  is given by 
$\pi_2(\Gamma;\Gamma_Q)=\frac{\theta}{\Gamma} \sin \theta=\left[\Gamma_Q\sqrt{(F_{\bot}-F_{/\!/})(F_{\bot}+F_{/\!/}-2\Gamma/\Gamma_Q)}\:\right]^{-1}$ for $F_{/\!/}<\Gamma/\Gamma_Q<(F_{\bot}+F_{/\!/})/2$ and $\pi_2(\Gamma;\Gamma_Q)=0$
 otherwise. The density of probability $\pi(\Gamma)$ for a QD to show a decay $\Gamma$ can be expressed as the probability $\pi_2(\Gamma;\Gamma_Q)$ of showing a decay $\Gamma$ for a given $\Gamma_Q$, summed over all possible values of $\Gamma_Q$ weighted by their probability $\pi_1(\Gamma_Q)$,:
 \begin{equation}
\pi(\Gamma) = \int_{\Gamma_Q}\pi_2(\Gamma;\Gamma_Q)\pi_1(\Gamma_Q)\mathrm{d}\Gamma_Q 
\end{equation}
The emission decay curve can finally be expressed as 
$I(t)=I_0\int_{\Gamma=0}^{+\infty}\int_{\Gamma_Q}\pi_2(\Gamma;\Gamma_Q)\pi_1(\Gamma_Q)\mathrm{d}\Gamma_Q\exp(-\Gamma t)\mathrm{d}\Gamma$. 
By fitting the experimental decay curves with this expression of $I(t)$, the free parameters $F_{\bot}$ and $F_{/\!/}$  are obtained.
For patch antenna 1, with diameter of 1.6~$\mathrm{\mu}$m, Purcell factors $F_{\bot}$=35 and $F_{/\!/}$=5 are deduced. For other antennas, values as large as 60 to 80 for $F_{\bot}$ are deduced while $F_{/\!/}$ is found to be in the 1-3 range showing that our observations are in excellent agreement with theoretical simulations. The minimum value of 20 measured for $F_{\bot}$ is lower than the smallest theoretical value probably due to deviation from the ideal circular shape of the antennas. Note that the above expressions are only valid for emitters centered in the antenna. However numerical calculations show that for QDs with an off-centering below 100 nm present the same Purcell factors as centered QDs.

\begin{figure}
 \includegraphics[width=6cm]{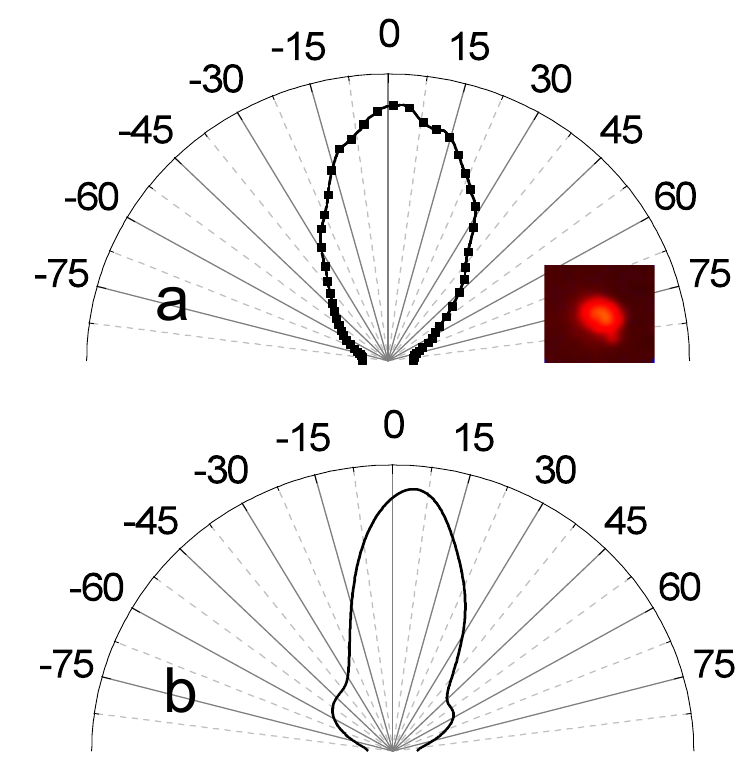}%
 \caption{Controlling the radiation pattern. a: Radiation pattern measured for a single cluster coupled to  patch antenna 1 with diameter 1.6~$\mathrm{\mu}$m and Purcell factors $F_{\bot}$=35 and $F_{/\!/}$=5. b: Calculated radiation pattern considering a cluster of randomly orientated QDs uniformly spaced in cylinder cluster with 10~nm height, 50~nm diameter, and located 15~nm from the disk center. Insert: Fluorescence image of the patch antenna 1.  \label{Fig4}}
 \end{figure}

We now study the modification of the angular far-field-emission pattern by Fourier imaging technique. We use a x100, NA 1.4 immersion oil objective to collect fluorescence light. The emission pattern of the antenna was deduced from the objective's back focal plane image observed on a high-sensitivity camera. Figure~\ref{Fig4}(a) presents the radiation pattern for the patch antenna 1. The radiation pattern features a nicely directional single lobe with an angular width of 35$^o$, which can be easily coupled to an optical fiber for long distance propagation. While calculated radiation patterns for a point emitter present several highly directional lobes, the finite size of the cluster leads to an averaging of the calculated radiation patterns on the whole cluster size. We model the experimentally observed radiation pattern considering emitters uniformly spaced in a cylinder cluster with 10~nm height. Using the cluster diameter and its position as the only fitting parameters, we find that the measured radiation pattern corresponds to the emission of a cluster with 50~nm radius, 15~nm away from the antenna gold disk center. This value is fully consistent with the typical cluster size measured using AFM (figure~\ref{Fig2}(c-d)).

Our work shows that optical patch antennas can control both the emission rate and the radiation pattern of quantum emitters. We have shown that large Purcell factors and high directionality can be obtained on micron size antennas with an accuracy for the emitter position of 25~nm. 
The acceleration of emission results from the coupling to the confined plasmon modes, which in turn can be radiatively or non radiatively dissipated. This first demonstration has been implemented with QDs emitting visible light, for which growth is well controlled and emission can be efficiently measured with standard detectors. While dissipation of plasmons is the dominant term in the visible range, we stress out that the same geometry and the same technology to position the emitters in the patch antenna can be directly transferred for emitters in the telecom wavelength like InAsP quantum dots in InP nanowire\cite{Borgstrom2005} or carbon nanotubes\cite{Hogele2008}.

\begin{acknowledgments}
This work was partially supported by the French ANR P3N DELIGHT, the ERC starting grant 277885 QD-CQED, the C'nano Ile-de-France NanoPlasmAA and the RTRA Triangle de la Physique (Project PAO). FB acknowledges the support of the french Direction Generale de l'Armement, BH acknowledges support from Cnano Ile de France.
Correspondence should be addressed to pascale.senellart@lpn.cnrs.fr, agnes.maitre@insp.jussieu.fr or jean-jacques.greffet@u-psud.fr.
\end{acknowledgments}


\end{document}